\documentclass[12pt]{iopart}
\usepackage{latexsym}
\usepackage{graphicx}
\usepackage{epsfig}
\usepackage[FIGTOPCAP]{subfigure}
\usepackage{amsfonts}
\usepackage{amssymb}
\usepackage{footmisc}
\usepackage{textcomp}

\begin{document}

\title{Focusing a deterministic single-ion beam}

\author{Wolfgang Schnitzler$^1$\footnote{corresponding author:
wolfgang.schnitzler@uni-ulm.de}, Georg Jacob$^1$, Robert Fickler$^2$, Ferdinand Schmidt-Kaler$^1$ and Kilian Singer$^1$}

\address{$^1$ Universit\"at Ulm, Institut f\"ur Quanteninformationsverarbeitung, Albert-Einstein-Allee 11, D-89069 Ulm, Germany}
\address{$^2$ Universit\"at Wien, Institut f\"ur Quantenoptik, Quantennanophysik \& Quanteninformation, Boltzmanngasse 5, A-1090 Wien, Austria}

\begin{abstract}

We focus down an ion beam consisting of single $^{40}\mathrm{Ca}^+$ ions to a spot size of a few \textmu m using an einzel-lens. Starting from a segmented linear Paul trap, we have implemented a procedure which allows us to deterministically load a predetermined number of ions by using the potential shaping capabilities of our segmented ion trap. For single-ion loading, an efficiency of 96.7(7)\,\% has been achieved. These ions are then deterministically extracted out of the trap and focused down to a 1$\sigma$-spot radius of $(4.6\pm1.3)$\,\textmu m at a distance of 257\,mm from the trap center. Compared to former measurements without ion optics, the einzel-lens is focusing down the single-ion beam by a factor of 12. Due to the small beam divergence and narrow velocity distribution of our ion source, chromatic and spherical aberration at the einzel-lens is vastly reduced, presenting a promising starting point for focusing single ions on their way to a substrate.

\end{abstract}

\pacs{03.67.-a; 29.25.Ni; 41.; 61.72.Ji; 85.40.Ry}


\newpage

\tableofcontents

\newpage
\section{Motivation} In the age of electronic data processing, the demand for more powerful and faster computers is paramount. As already stated by Gordon E. Moore in 1965, the number of transistors that can be placed inexpensively on an integrated circuit has increased exponentially, doubling approximately every two years. The miniaturization of semiconductor devices has reached length scales of a few tens of nanometers where statistical Poissonian fluctuations of the number of doping atoms in a single transistor significantly affect the characteristic properties of the device, e.g. gate voltage or current amplification. Especially when thinking about future solid state quantum processors, statistical fluctuations of the dopant concentration are fatal for systems, which are based on single implanted qubit carriers like color centers in diamond or phosporous dopants in silicon  \cite{GURUDEV2007,NEUMANN2008,KANE1998,GREENTREE2008}. Until recently, most implantation techniques resorted to thermal sources, where a control of the number of doping atoms is only possible via a postdetection of the implantation event. However, although a wide range of postdetection schemes (e.g. the observation of Auger electrons, the generation of electron-hole pairs or changes in the conductance of field effect transistors) is available \cite{SHINADA2002,PERSAUD2004,MITIC2005,BATRA2007,SHINADA2008}, most of these techniques either require highly charged ions or high implantation energies which, as a down side, generate unintentional defects in the host material. Another fabrication method revolves around the structuring of chemically treated Si-surfaces. Using a hydrogen terminated Si-surface as starting point, the tip of a tunneling microscope allows for removing single hydrogen atoms which are then replaced by doping atoms such as phosphorus due to a chemical reactive surface binding \cite{OBRIEN2001,SCHOFIELD2003,RUESS2004,POK2007,RUESS2007}. Although this technique allows for placing single dopants with sub-nm resolution, the applicability is mainly restricted to specific substrates. In order to circumvent the necessity of any post-detection schemes and to expand the applicability to a wider range of elements, deterministic single-ion sources on the basis of optical lattices \cite{GREINER2007}, magneto-optical \cite{MCCLELLAND2003B,HANSSEN2006,HANSSEN2008} and segmented linear Paul traps \cite{MEIJER2006} have been developed.

Here, we present the experimental setup to focus down an ion beam consisting of single $^{40}\mathrm{Ca}^+$ ions to a spot size of a few \textmu m by utilizing an einzel-lens. The single-ion beam is generated by a segmented linear Paul trap which allows us to deterministically extract single ions on demand \cite{MEIJER2006,SCHNITZLER2009,MEIJER2008}. Due to the small beam divergence and narrow velocity distribution of our single-ion source, chromatic and spherical aberration at the einzel-lens is strongly reduced presenting a promising starting point for focusing single ions onto a prospective substrate.

The paper is organized as follows: We first describe the experimental setup which is used for trapping, imaging and manipulation of the $^{40}$Ca$^{+}$ in section~\ref{ExperimentalSetup} including a detailed description of the ultra high vacuum setup, the specially designed segmented linear ion trap, the optical setup as well as the extraction mechanism and the utilized ion optics. In section~\ref{ExperimentalResults}, we present the experimental results, namely the deterministic loading of single $^{40}\mathrm{Ca}^+$ ions and the focusing of the single-ion beam into a 5\,\textmu m spot. Finally, in section \ref{Conclusion}, we give a short conclusion and sketch possible future applications.

\section{Experimental setup}
\label{ExperimentalSetup}

\subsection{Ultra high vacuum setup}
\label{VacuumSetup}

The core piece of the ultra high vacuum (UHV) setup is a Magdeburg hemisphere from Kimball Physics\footnote{Kimball Physics Inc., USA} with an inner diameter of 91.5\,mm and a total length of 106\,mm. It features a total of seven different access ports with flange sizes ranging from DN16CF up to DN63CF. A specifically designed segmented linear ion trap (see chapter~\ref{RailTrap}) is mounted inside of the hemisphere with the trap axis being located 165\,mm above the optical table.

\begin{figure}[htb]
\begin{center}
\includegraphics[width=13.5cm]{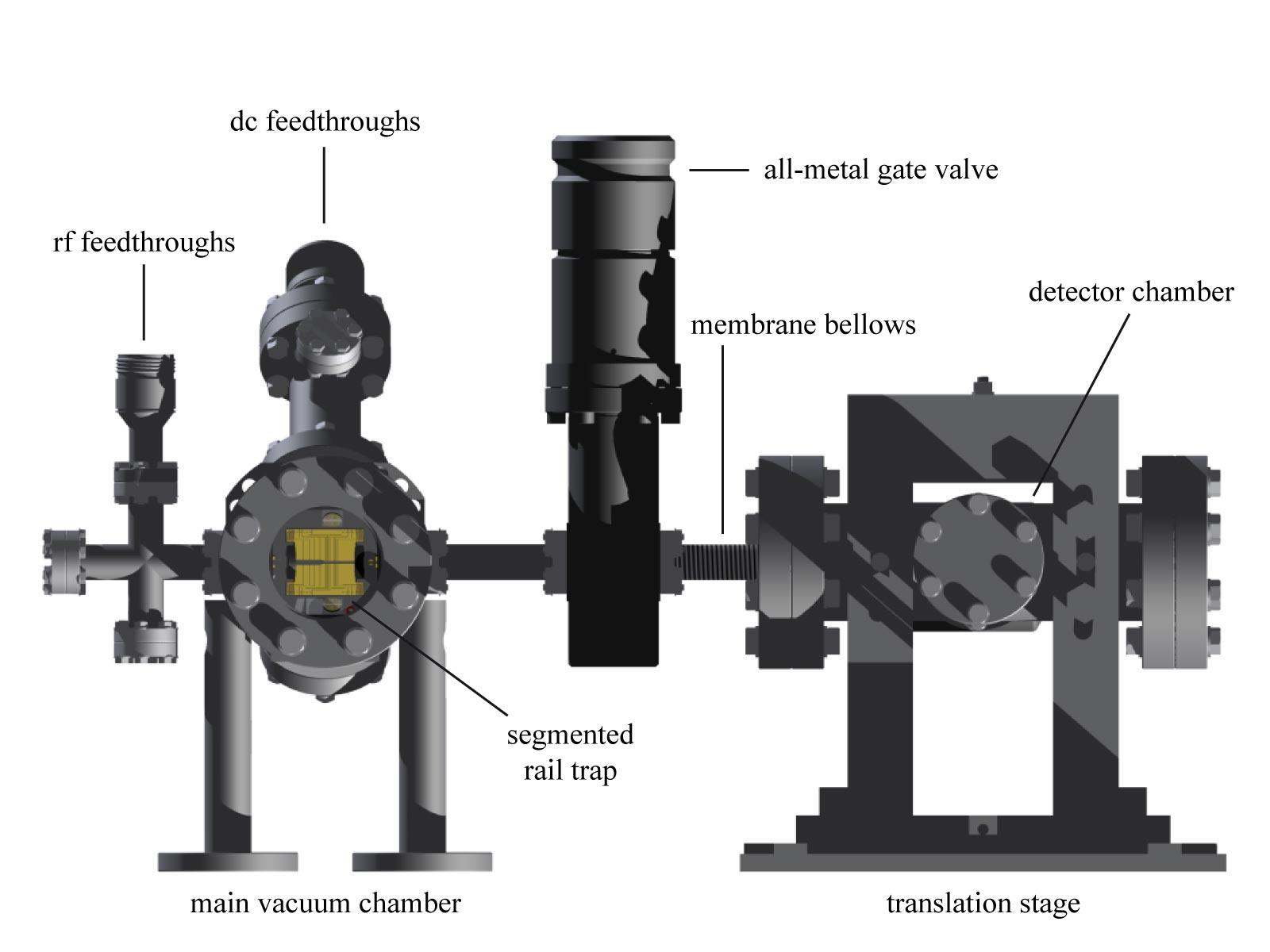}
\caption{Schematic drawing of the UHV setup. The setup consists of a main vacuum chamber and a detector chamber which are connected via an all-metal gate valve. A membrane bellows in combination with a specifically designed translation stage allows for rotating the detector chamber in horizontal and vertical direction with respect to the trapping position of the ion.}
\label{fig:VacuumSetup}
\end{center}
\end{figure}

All access ports are equipped with fused silica (SiO$_2$) windows allowing for optical access with laser beams and for imaging the ions.  A 20-pin vacuum feedthrough for the dc electrodes and the ovens is connected via a T-piece to the hemisphere whereas the power supply for the rf electrodes is managed using a separate 4-pin vacuum feedthrough which is mounted on a four-way cross, see also Fig.~\ref{fig:VacuumSetup}. Using an all-metal gate valve from VAT\footnote{VAT Deutschland GmbH, Germany}, the main vacuum chamber is connected to a separate detector chamber which is attached to a specifically designed translation stage. In combination with a membrane bellows, the translation stage allows for rotating the detector chamber in horizontal and vertical direction with respect to the trapping position of the ion compensating any misalignment between the trap axis and the optical axis of the utilized ion optics. The complete setup measures 850\,mm in length with a depth of 480\,mm and a total height of about 380\,mm. The pressure is held by a 20\,l/s VacIon \textit{Plus} StarCell ion getter pump and a titanium sublimation pump\footnote{Varian Inc., USA}. In addition, both the main vacuum and the detector chamber are connected via all-metal angle valves to a turbomolecular pump from Oerlikon\footnote{Oerlikon Leybold Vacuum GmbH, Germany}. The base pressure of the UHV setup equals 5$\times10^{-10}$\,mbar which is measured using a Varian UHV-24p Bayard-Alpert type ionization gauge tube.

\subsection{Segmented linear rail trap}
\label{RailTrap}

\begin{figure}[htb]
\begin{center}
\includegraphics[width=11.5cm]{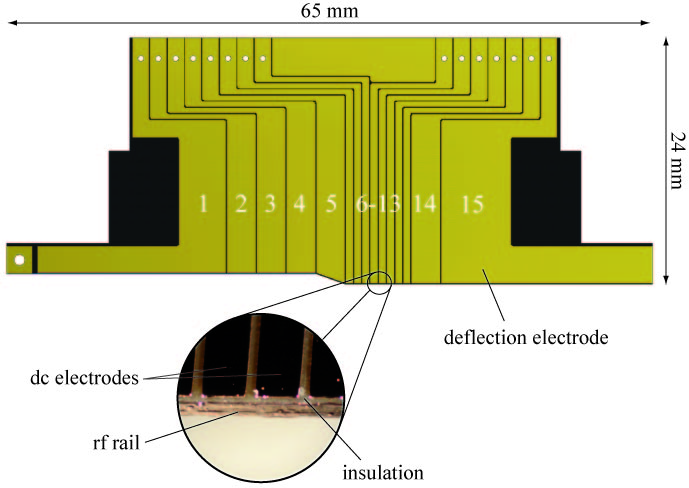}
\caption{Technical drawing of one of the four trap blades featuring a total of 15 independent dc electrodes. A wide loading zone (electrodes 1-4) is connected via a taper (electrode 5) to a narrow experimental zone (electrodes 6-14). A deflection electrode (electrode 15) is used to alter the trajectories of the extracted ions. The rf rail has a thickness of about 22\,\textmu m and covers the whole front of the blade. Isolated parts are colored black. The close-up view shows a microscope image of the front part of the blade.}
\label{fig:Trap}
\end{center}
\end{figure}

The trap consists of four copper plated polyimide blades which were manufactured using printed circuit board (PCB) technology and are arranged in a x-shaped manner \cite{HUBER2008}. It features a total of 15 independent dc electrodes which can be assigned to three different trap sections: A wide loading zone (electrodes 1-4) is connected via a taper (electrode 5) to a narrow experimental zone (electrodes 6-14). A deflection electrode (electrode 15) is used to alter the trajectories of ions which are extracted out of the trap. In order to generate the radial confinement, an additional electrode is running along the inner front face of each blade which will be referred to as rail in the following. The blade itself has a total length of 65\,mm and a thickness of 410\,\textmu m with a 18\,\textmu m copper plating on both sides. Electrodes 2-5 as well as electrode 14 have a width of 2.8\,mm, electrodes 6-13 have a width of 0.7\,mm, respectively. In order to efficiently deflect ions by targeted application of low voltages, the deflection electrode has been elongated in extraction direction featuring a total length of 20.6\,mm. The distance between the rf rail and the dc electrodes as well as between adjacent dc electrodes equals 0.1\,mm, see also Fig.~\ref{fig:Trap} for a technical drawing of one of the four trap blades.

The distance between the inner front faces of opposing blades equals 4\,mm in the loading zone and 2\,mm in the experimental zone, respectively. In case of a standard Paul trap, the rf voltage is supplied to two of the four blades. The other two blades are used for the axial confinement and are therefore divided into several electrodes. However, in the case of our rail trap all four blades are identical. Apart from the deflection electrodes, corresponding electrodes on all four blades (i.e. electrode 1 of each of the blades) are electrically connected together resulting in a stronger and more symmetrical axial confinement than in the standard case. The rf voltage is only applied to the rails of two opposing blades; the front faces of both other blades are grounded generating the quadrupole potential for the radial confinement. As these rails are running along the whole trap axis and are continuing round the corner at the end of each blade, the radial confinement is sustained during the whole extraction process and a targeted shooting is accomplished.

Under typical operating conditions, we apply a voltage of 400\,V$_\mathrm{pp}$ at the frequency of $\Omega/2\pi$ = 12.155\,MHz to the rf electrodes leading to a radial secular frequency $\omega_\mathrm{rad}/2\pi$ = 430\,kHz for a $^{40}\mathrm{Ca}^+$ ion. The dc-electrode trap segments 7 and 13 are supplied with 35\,V and the remaining electrodes with 0\,V resulting in an axial trapping potential with $\omega_\mathrm{ax}/2\pi$ = 280\,kHz. The location of trapped ions is above electrode 10.

\subsection{Optical setup}
\label{OpticalSetup}

The generation of the necessary laser beams is mainly done by using commercial grating stabilized diode laser systems from TOPTICA\footnote{TOPTICA Photonics AG, Germany}. The ions are illuminated by resonant laser light near 397\,nm and 866\,nm for Doppler cooling. Scattered photons are collected by a f/1.76 lens on an EMCCD camera\footnote{Andor Technology, model DV885KCS-VP} to image individual $^{40}\mathrm{Ca}^+$ ions. From the width of the laser excitation spectrum on the $S_\mathrm{1/2}$ - $P_\mathrm{1/2}$ laser cooling transition, we deduce a temperature of about 2\,mK slightly above the Doppler cooling limit. Calcium and dopant ions are generated in a multi-photon ionization process by a pulsed frequency tripled Nd-YAG laser at 355\,nm with a pulse power of 7\,mJ. Dopant ions are sympathetically cooled and identified from the voids in the fluorescence image compared to that of a regular linear $^{40}\mathrm{Ca}^+$ crystal. The species of dark ions can be identified by exciting collective vibrational modes with an ac voltage applied to electrode 9 and observing a blurring of the $^{40}\mathrm{Ca}^+$ fluorescence image at the resonance frequency $\omega_\mathrm{ax.}$ \cite{NAEGERL1998}. Alternatively, amplitude modulated resonant laser light is used \cite{DREWSEN2004} to determine the charge to mass ratio of trapped particles at a precision of better than 0.2\,\%. In order to suppress vibrations of the building and to reduce thermal elongations to a minimum, the complete setup is located on an air-suspended optical table which is situated in a temperature stabilized laboratory.

\subsection{Extraction mechanism and setup}
\label{ExtractionSetup}

Initially, the ion is trapped at the center of electrode 10 by supplying a voltage of 35\,V to electrodes 7 and 13, see also Fig.~\ref{fig:AxialPot}(a) for a simulation of the trapping potential in axial direction. However, due to the distance from the electrodes to the trap center, the generated axial trapping potential only features a depth of about 4\,eV not representing a large barrier during the extraction process. The extraction process is then initiated by increasing the dc voltages on electrodes 9 and 10 to 500\,V within a few tens of nanoseconds. The extraction voltage generates a potential hill at the position of the ion effectively canceling the axial confinement. However, as the ion already gets accelerated by the developing potential during the switching process, the ion doesn't sense the full potential strength and the effective energy transfer is lessened. In addition, due to the asymmetric voltage configuration the peak voltage is not at the position of the ion and it is only accelerated by the shoulder of the generated potential hill reducing the kinetic energy even further. From a time-of-flight analysis, one can deduce a final kinetic energy of about 80\,eV for the extracted $^{40}\mathrm{Ca}^+$ ions \cite{SCHNITZLER2009}.

\begin{figure}[htb]
\begin{center}
\includegraphics[width=11cm]{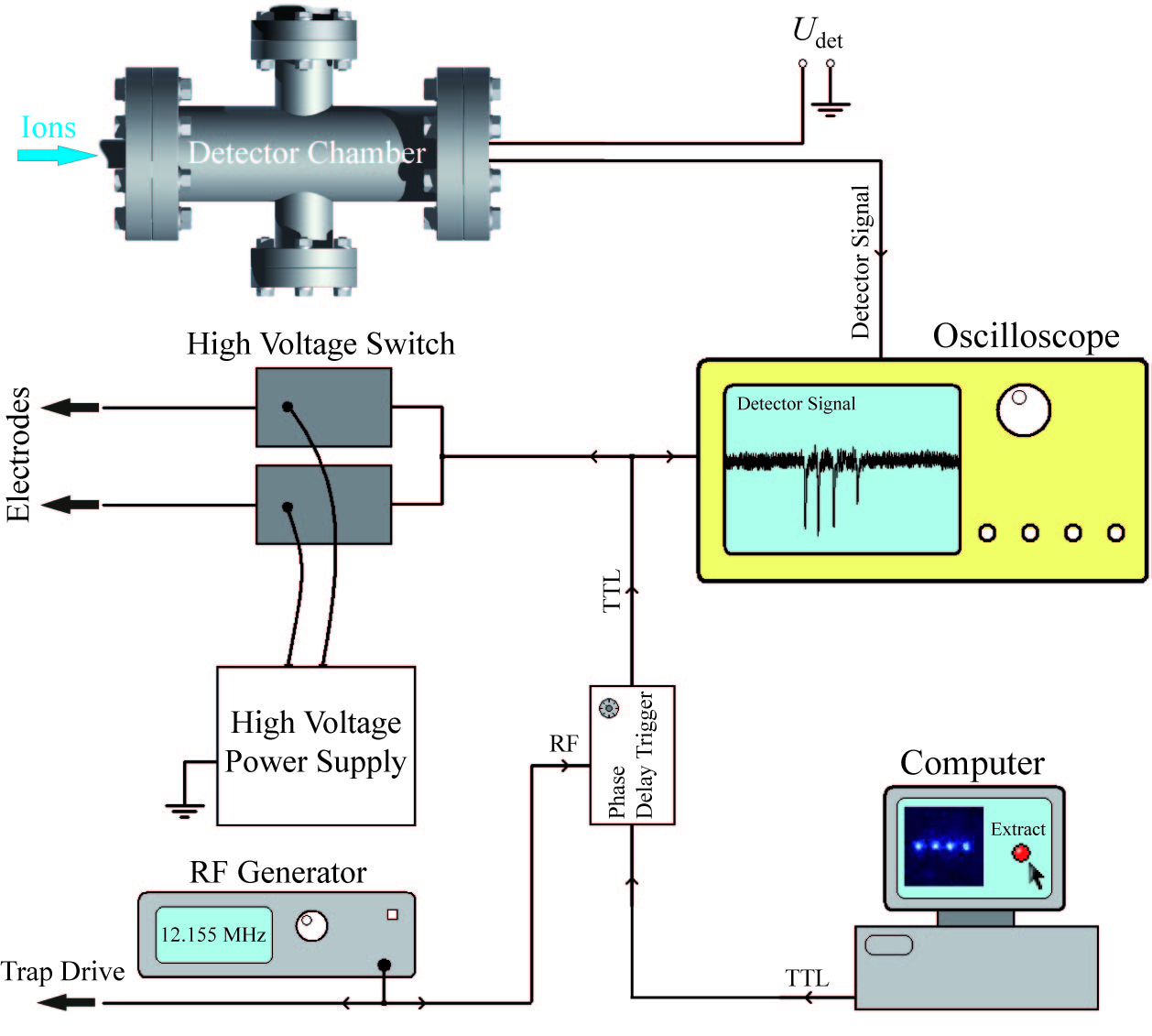}
\caption{Schematic drawing of the extraction setup. After a certain number of ions has been loaded into the trap, the extraction process is triggered via a TTL signal which is synchronized to the trap drive frequency $\Omega$ using a phase synchronization circuit (phase delay trigger). Two high voltage switches then increase the dc voltages on electrodes 9 and 10 to 500\,V within a few tens of nanoseconds leading to the extraction of the trapped ions. Upon entering the detector chamber, the extracted ions are detected using an electron multiplier tube leading to signal dips on an oscilloscope.}
\label{fig:ExtractionSetup}
\end{center}
\end{figure}

Fig.~\ref{fig:ExtractionSetup} shows a schematic drawing of the extraction setup. After a specified number of ions has been loaded into the trap (in this case a linear ion crystal consisting of four $^{40}$Ca$^{+}$ ions), the extraction process is triggered via a computer-controlled TTL signal which is fed in a phase synchronization circuit (phase delay trigger). The phase synchronization circuit delays the TTL signal such that a constant delay to the next zero crossing of the trap drive frequency $\Omega$ is ensured as the synchronization is crucial in order to minimize shot to shot fluctuations of the velocity and position of the extracted ions. The measured delay time shows a 1$\sigma$-spread of 0.34\,ns. The switching of the extraction voltage\footnote{isec inc., model EHQ-8010p} is experimentally realized by two high voltage switches\footnote{Behlke inc., model HTS 41-06-GSM} which can switch voltages of up to 4\,kV within a timespan of 20\,ns. However, as the extraction voltage leads to a temporary charging of the trap, the TTL trigger signal is only supplied for a few ms to the high voltage switches reducing the unintentional charging to a minimum.

The detection of the extracted ions is then performed via an electron multiplier tube (EMT)\footnote{ETP inc., model AF553} with 20 dynodes and a main entrance aperture of 20\,mm which can detect positively charged ions with a specified quantum efficiency of about 80\,\% and a specified gain of 5$\times10^5$. Under typical operating conditions, the detector is supplied with a voltage of -2.5\,kV leading to detection signals with a width of 10 to 15\,ns and an amplitude of about 100\,mV which are recorded via an oscilloscope\footnote{Agilent infiniium 54832D MSO}. As already stated in chapter~\ref{VacuumSetup}, the detector is housed in a separate vacuum chamber with a distance of 287\,mm from the trap center.

\subsection{Ion optics}
\label{IonOptics}

\begin{figure}[htb]
\begin{center}
\includegraphics[width=14.5cm]{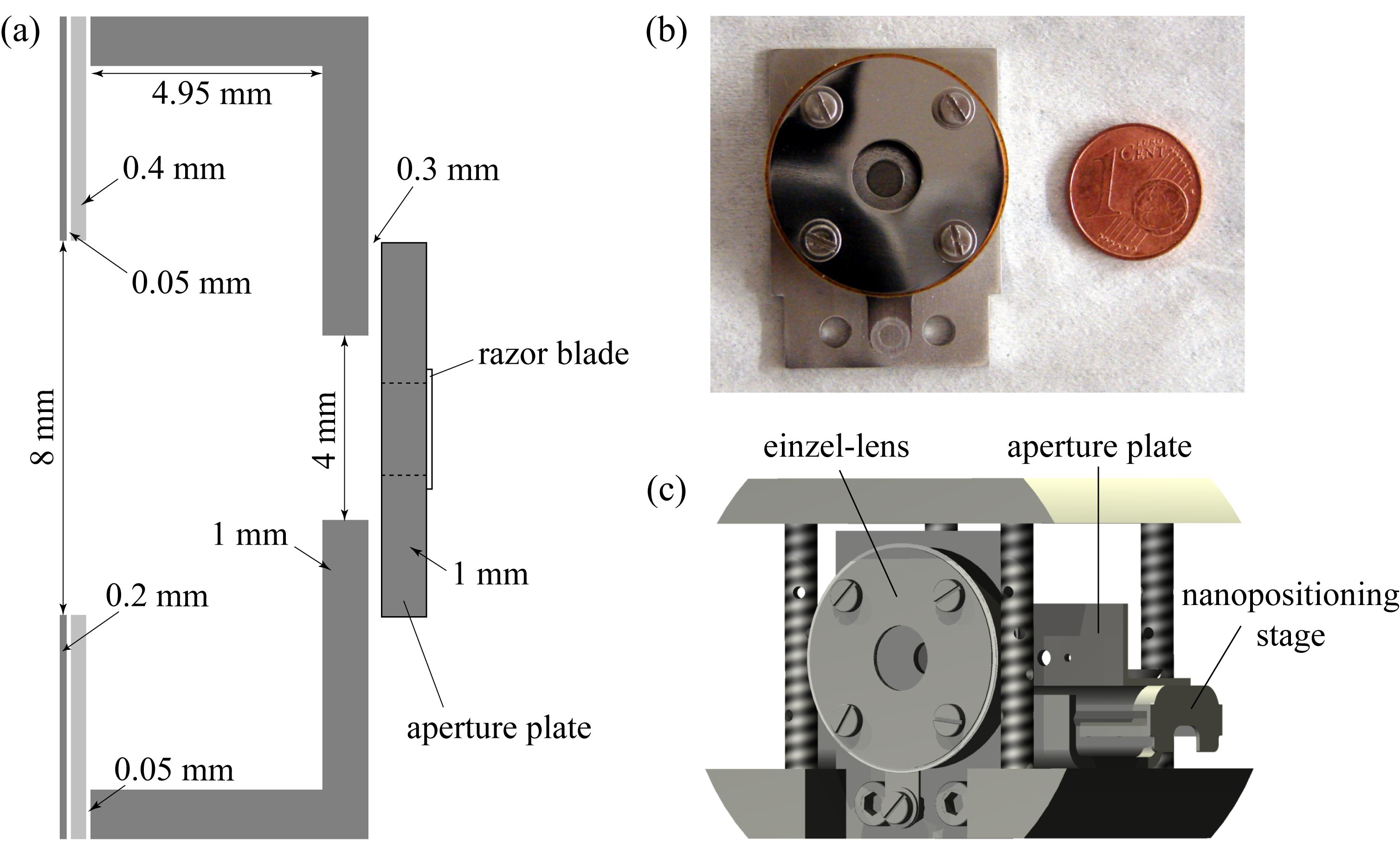}
\caption{(a) Schematic drawing of the einzel-lens with the dc electrode depicted in light grey and grounded electrodes colored dark grey, respectively. The razor blade (depicted in white), which is mounted behind the 2\,mm aperture of the movable aperture plate, is also included. (b) Picture of the assembled einzel-lens with an one-cent coin for size comparison. Under typical operating conditions, the dc electrode is supplied with a voltage of 150\,V. (c) Schematic drawing of the einzel-lens and the nano positioning stage which are located at the beginning of the detector chamber directly in front of the EMT.}
\label{fig:Lens}
\end{center}
\end{figure}

The utilized einzel-lens consists of three stainless steel electrodes which feature an outer diameter of 27\,mm. The first electrode has a thickness of 0.2\,mm with an inner diameter of 8\,mm, the second electrode has a thickness of 0.4\,mm and an inner diameter of 8\,mm, respectively. The third electrode consists of two parts: The first part is a cylindrical shaped spacer with an inner diameter of 16\,mm and a thickness of 4.95\,mm whereas the second part consists of a stainless steel plate with a thickness of 1\,mm featuring a 4\,mm aperture which is also used as mounting for all other electrodes. The first and third electrode are conductively interconnected whereas the second electrode is electrically insulated by using Kapton sheets with a thickness of 0.05\,mm, see also Fig.~\ref{fig:Lens}(a) for a schematic drawing of the einzel-lens. The calculation of the electrostatic potentials of the einzel-lens and the ion ray tracing simulations are performed by a high accuracy boundary-element solver package \cite{SINGER2009}. A detailed description of the imaging characteristics for different types of lens configurations can be found in \cite{FICKLER2009}.

All parts were electropolished in order to remove ridges which were formed during the manufacturing process and to reduce the overall surface roughness. For electropolishing we used a solution of phosphoric acid (H$_3$PO$_4$), methanesulfonic acid (CH$_3$SO$_3$H) and triethanolamine (C$_6$H$_{15}$NO$_3$) \footnote{patent specification DE102006050317B3}. The parts were polished for a timespan of 15\,min at a temperature of $65\,^{\circ}{\rm C}$ using an anodic current density of 5\,A/dm$^2$. Fig.~\ref{fig:Lens}(b) shows a picture of the assembled einzel-lens with an one-cent coin for size comparison. Under typical operating conditions, the second electrode of the einzel-lens is supplied with a voltage of 150\,V whereas the other two electrodes are grounded.

In order to determine the diameter of the single-ion beam, a movable aperture plate was installed between the einzel-lens and the detector featuring hole diameters ranging from 5\,mm down to 300\,\textmu m. In addition, a razor blade was attached behind the 2\,mm aperture effectively creating a well defined tearing edge (see also Fig.~\ref{fig:Lens}(a) and chapter~\ref{SpotSize}). The plate is mounted on a nano positioning stage\footnote{SmarAct, model SL-2040}, which allows for moving the aperture plate with an accuracy of a few tens of nanometers. An internal sensor module guarantees that the absolute position of the stage is kept constant. Fig.~\ref{fig:Lens}(c) shows a schematic drawing of the einzel-lens and the nano positioning stage which are located at the beginning of the detector chamber directly in front of the EMT.

\section{Experimental results}
\label{ExperimentalResults}

\subsection{Deterministic single-ion loading}

In order to reduce the divergence of the ion beam to a minimum, it is necessary to ensure that only one ion is extracted out of the trap with each shot. In case of a Paul trap, there are in principal two options to achieve this goal: The first option is to avoid the loading of more than one ion at a time. Since this is not feasible due to the fact that the number of loaded ions follows a Poissonian distribution for equal loading times, one needs to implement a feedback control in order to stop the ionization after one ion has been loaded into the trap. As it turned out, it was not possible to implement this method in our setup as the time which elapses between the ionization and the crystallization of an one-ion crystal has to be much shorter than the average time between ionization incidents of two different ions. Therefore, we have chosen the second possibility, namely to get rid of the redundant ions after the crystallization process. For this task, we have come up with a very simple and robust procedure which will be described in the following.

\begin{figure}[htb]
\begin{center}
\subfigure[]{
	\includegraphics[width=7.5cm]{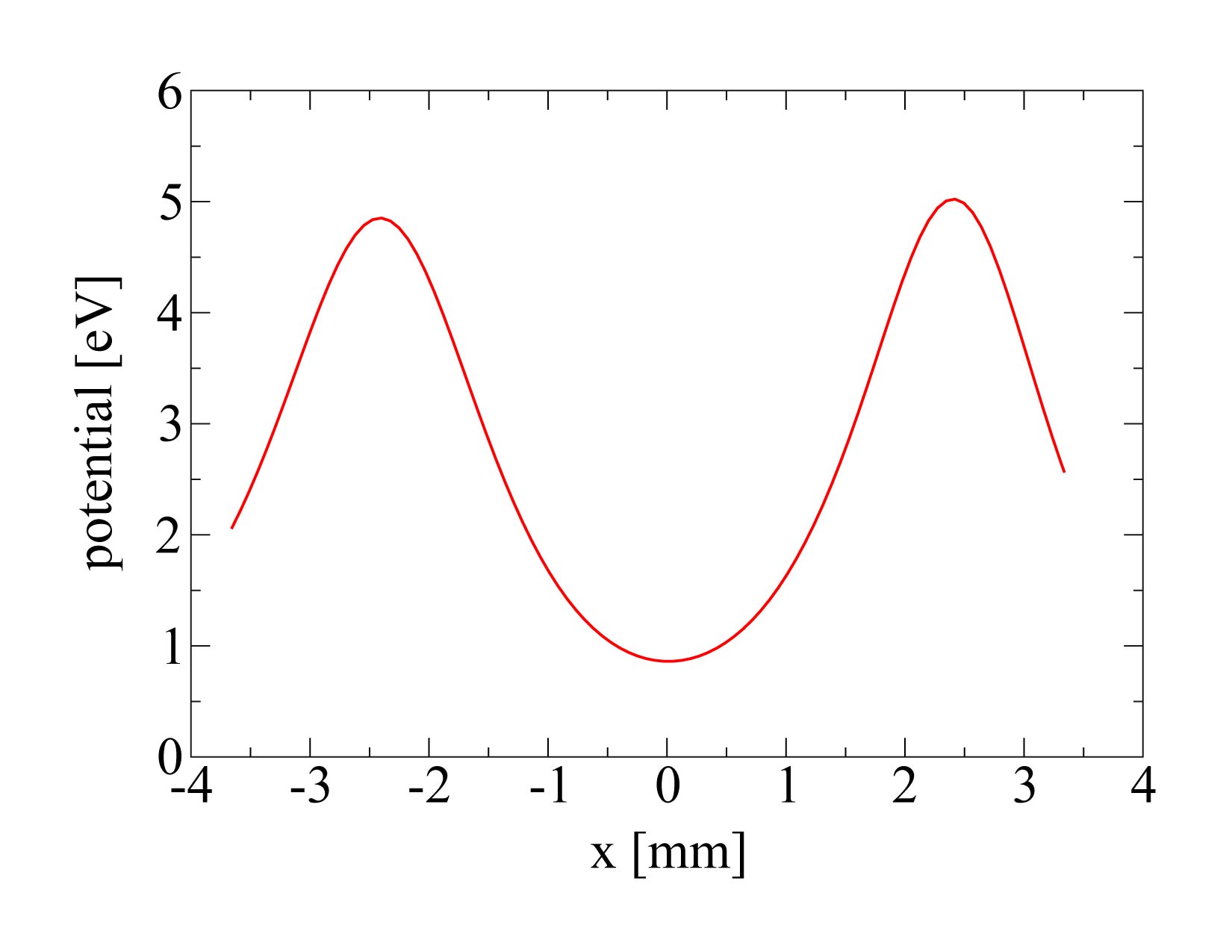}
}
\subfigure[]{
	\includegraphics[width=7.5cm]{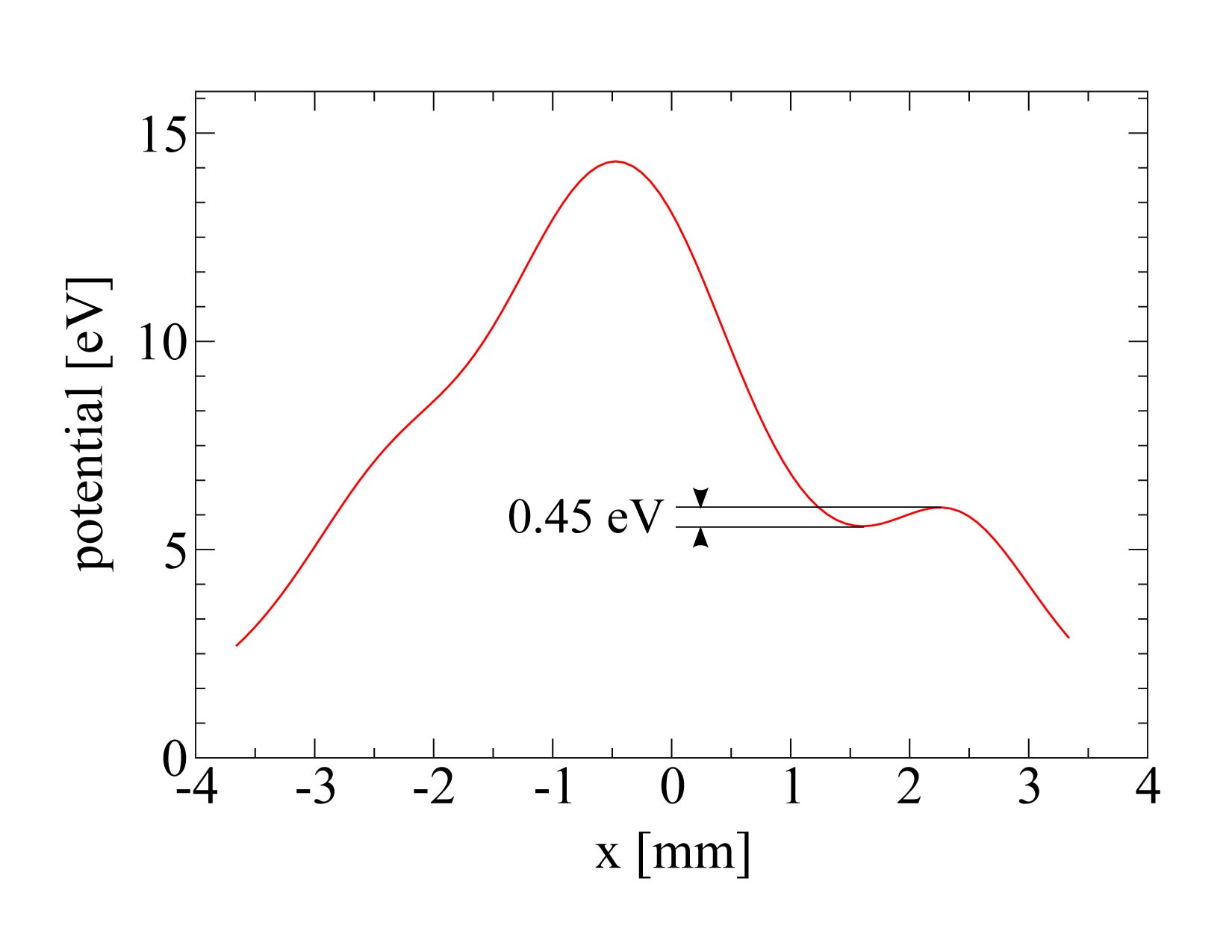}
}
\caption{(a) Simulated trapping potential in axial direction. The potential is generated by supplying dc electrodes 7 and 13 with a voltage of 35\,V and all remaining electrodes with 0\,V resulting in an axial trapping frequency $\omega_\mathrm{ax.}$ = $2\pi\times$280\,kHz and a trap depth of about 4\,eV. The center of electrode 10 is used as the point of reference for the axial position. (b) Same potential as in (a) but with a voltage of 54.8\,V applied to electrodes 9 and 10 reducing the axial trap depth to about 0.45\,eV in order to get rid of the surplus ions.}
\label{fig:AxialPot}
\end{center}
\end{figure}

To drop the surplus ions, we exploit the potential shaping capabilities offered by our segmented linear rail trap (see also chapter~\ref{RailTrap}). At the beginning, the ions are trapped above electrode 10 by supplying a voltage of 35\,V to electrodes 7 and 13 leading to a axial trapping potential with a trap depth of about 4\,eV, see also Fig.~\ref{fig:AxialPot}(a). By using the offset input of our high voltage switches (see also chapter~\ref{ExtractionSetup}), the voltage supplied to electrodes 9 and 10 is linearly ramped up from 0 to 54.8\,V within a timespan of about 50\,ms before it is again linearly reduced to 0\,V after a waiting time of 200\,ms. As a consequence, the depth of the axial trapping potential is reduced to about 0.45\,eV thus becoming shallow enough that only one ion will remain inside of the trap, see also Fig.~\ref{fig:AxialPot}(b). In order to determine the reproducibility of our method, we repeated the procedure described above with ion crystals of various sizes yielding a total efficiency of 96.7(7)\,\% for reducing the amount of ions to exactly one. By means of this approach, it is also possible to reduce the number of ions in the chain to an arbitrary predefined quantity, e.g. from 5 to 4 ions, but since this is not used for the experiments below, we do not discuss it any further.

\subsection{Experimental determination of the spot size}
\label{SpotSize}

In former measurements, we determined the size of our ion beam by shooting single ions through the 300\,\textmu m aperture, yielding a spot radius of 83$(^{+8}_{-3})$\,\textmu m and a full-angle beam divergence of 600\,\textmu rad at a distance of 257\,mm from the trap center \cite{SCHNITZLER2009}. Here, 68.3\,\% of all detected ions were located inside of the stated spot radius. However, in order to compare these values with our current measurements, we converted them according to equation~(\ref{eq:SigmaSpotRadius}), resulting in a 1$\sigma$-spot radius of 55$(^{+5}_{-2})$\,\textmu m. To further improve the resolution of our single-ion beam, we installed an einzel-lens between the trap and the detector, see also chapter~\ref{IonOptics}. In order to obtain an optimal alignment with respect to the optical axis of the einzel-lens, the radial adjustment of the ion beam is performed by utilizing the deflection electrodes of our segmented rail trap. Using the 1\,mm aperture behind the einzel-lens, the optimal deflection voltages are retrieved by gradually changing them, maximizing the hit rate on the detector. As it turns out, the hit rate is very sensitive to changes of the deflection and compensation voltages, i.e. a minor change in the range of a few mV is already sufficient to miss the 1\,mm aperture effectively reducing the measured hit rate to zero. Hence, to keep adjacent measurements consistent to each other, all measurements presented were conducted with a fixed set of deflection and compensation voltages.

In order to determine the spot size of the ion beam in the focal plane of the installed ion optics, we utilize the aforementioned razor blade which is mounted behind the 2\,mm aperture, see also chapter~\ref{IonOptics}. By stepwise moving the razor blade into the path of the generated single-ion beam, a gradual cut-off of the ion trajectories is achieved leading to a reduction of the overall hit rate on the detector. To derive the 1$\sigma$-spot radius, these points are then fitted with an error function which itself is obtained by integrating a Gaussian distribution which will be assumed for the radial cross-section of the single-ion beam:

\begin{equation}
f(x)=\int\limits_{-\infty }^{x} \frac{c}{\sigma\,\sqrt{2\pi}}\,
e^{-\frac{1}{2}\left(\frac{y-a}{\sigma}\right)^2}\,
\mathrm{d}y= \frac{c}{2} \left(1 +
\mathrm{erf}\left(\frac{x-a}{\sigma\,\sqrt{2}}\right)\right).
\label{eq:SigmaSpotRadius}
\end{equation}

Here, $\sigma$ denotes the radius of the ion beam with $a$ being the offset in x-direction. The parameter $c$ is used as a scaling factor in order to account for the detector efficiency which has been experimentally measured to be 0.87 \cite{SCHNITZLER2009}. As the time span needed for doing a single measurement is considerably long ($\approx3$ incidents per minute), we had to reduce the number of measurements taken for each blade position to a total of 10 individual shots. This was done in order to avoid changes of the deflection and compensation voltages which would have been necessary to counteract long term drifts caused by a thermal expansion of the trap.

\begin{figure}[htb]
\begin{center}
\includegraphics[width=12.5cm]{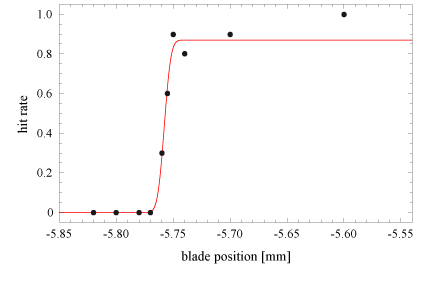}
\caption{Experimental determination of the spot size. The graph shows the hit rate for single ions as a function of the blade position for a lens voltage of 150\,V, resulting in a 1$\sigma$-spot radius of $(4.6\pm1.3)$\,\textmu m.}
\label{fig:Conclusion}
\end{center}
\end{figure}

As the exact focal distance of our einzel-lens is unknown, one would normally have to repeat the measurement described above for different spacings between the einzel-lens and the razor blade. However, as our nano positioning stage is only capable of moving perpendicular to the ion beam, the location of the focal plane of the einzel-lens has to be adjusted by changing the lens voltage. Measuring the spot radius for different lens voltages in the plane of the razor blade therefore should allow for obtaining the minimal spot radius. In the experiment, we have measured a 1$\sigma$-spot radius of $(4.6\pm1.3)$\,\textmu m for a lens voltage of 150\,V, see also Fig.~\ref{fig:Conclusion}. Compared to the original beam diameter, our einzel-lens is thus capable of focusing the single-ion beam by a factor of 12.

\begin{figure}[htb]
\begin{center}
\includegraphics[width=10cm]{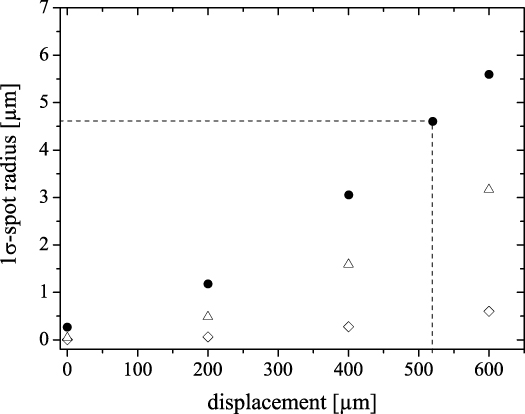}
\caption{Simulated 1$\sigma$-spot radii as a function of the spatial displacement of the ion beam in the principal plane of the lens. The radii were calculated for an initial ion temperature of 2\,mK with ($\bullet$) and without ($\vartriangle$) deflection voltages as well as for an initial ion temperature of 100\,\textmu K without deflection voltages ($\diamond$). In order to explain the experimentally measured 1$\sigma$-spot radius of 4.6\,\textmu m for 2\,mK laser-cooled ions, a displacement of half a millimeter is already sufficient (dashed lines). Note, that the initial ion temperature and deflection voltages for the simulated data set ($\bullet$) match the experimental conditions whereas the spatial displacement was not determined experimentally.}
\label{fig:SimulatedSpots}
\end{center}
\end{figure}

Numerical simulations predict that an even higher resolution should be feasible. In case of ground-state cooled ions with a temperature of 100\,\textmu K, simulations predict a 1$\sigma$-spot radius of 1\,nm, which is enlarged to 45\,nm for a temperature of 2\,mK. However, in the experiment, additional deflection voltages are needed in order to align the ion beam with respect to the optical axis of the utilized einzel-lens. Including these deflection voltages in the numerical simulations leads to a further decline in resolution, resulting in a 1$\sigma$-spot radius of about 270\,nm. Only taking into account a spatial displacement of the incident ion beam in the principal plane of the lens, numerical simulations predict that a mismatch of 520\,\textmu m with respect to the optical axis is already sufficient in order to obtain a spot size of 4.6\,\textmu m in the focal plane of the lens, see also Fig.~\ref{fig:SimulatedSpots}. In case of our experiment, this displacement is caused by a mechanical misalignment of the center of the aperture with respect to the optical axis of the einzel-lens as well as a displacement of the ion beam itself within the 1\,mm aperture. The former can be ascribed to manufacturing imperfections, which are estimated to be about 200\,\textmu m. The latter is caused by the initial adjustment of the ion beam using the 1\,mm aperture. By taking into account the original beam diameter, a displacement of less than 400\,\textmu m can be assumed. Hence, both errors result in a maximum total deviation of 600\,\textmu m, explaining the numerically predicted mismatch. A more compact experimental setup as compared with the total length of about 260\,mm (see Fig.~\ref{fig:VacuumSetup}) is envisioned for the future.

\section{Conclusion and Outlook}
\label{Conclusion}

Using a segmented linear Paul trap as a deterministic point source for laser-cooled ions, we have demonstrated the focusing of an ion beam consisting of single $^{40}\mathrm{Ca}^+$ ions to a spot size of a few \textmu m. By utilizing the potential shaping capabilities of our ion trap, we were able to deterministically load a predetermined number of ions which will allow for further automatization of the loading and extraction procedure resulting in orders of magnitude increased repetition rates. In future experiments, efforts will be made to avoid the usage of the deflection electrodes by aligning the beam mechanically and to implement more sophisticated cooling techniques like sideband or EIT cooling in order to reduce the divergence of the ion beam. Moreover, the utilized nano positioning stage will be exchanged by a modular setup consisting of three independent stages in order to move apertures and substrates in all three spatial directions. The additional degrees of freedom will not only allow for a more precise alignment with respect to the optical axis of the einzel-lens but will also allow us to determine the optimal focal length and spot size more accurately. Using two perpendicular edges of the substrate in the same way as the razor blade will then allow for approaching a specific location on the substrate relative to the corner formed by the two edges. Alternatively, one could think of mounting the einzel-lens on the tip of an AFM, thus allowing for determining the implantation position with respect to potential surface structures \cite{MEIJER2008}. A new lens design, which is currently in development, will facilitate a better focusing of the single-ion beam, and additionally allow for post-accelerating the extracted ions in order to reach higher implantation energies of a few keV, i.e.~improving the conversion efficiency for the generation of color centers in diamond \cite{JELEZKO2006}. Due to the high spatial and temporal resolution of the focused single-ion beam, one might also consider to shoot ions from one ion trap into another. This would allow to transport quantum information stored in the internal, electronic states of the ions in a much faster way and over larger distances than currently possible when performing a standard ion transport in multi-segmented micro ion traps.

\vspace{1cm}

\textbf{Acknowledgments:} We acknowledge financial support by the Landes\-stiftung Baden-W\"urttemberg in the framework 'atomics' (Contract No. PN 63.14) and the excellence programme.

\vspace{1cm}

\bibliographystyle{unsrt}
\bibliography{bibliography}

\end{document}